\newif\ifisarxiv
\isarxivtrue
\ifisarxiv
\documentclass[10pt]{article}
\else
\documentclass[twocolumn,10pt]{article}
\fi
\usepackage{fullpage}
\usepackage{amsmath}
\usepackage{amssymb}
\usepackage{amsthm}
\usepackage{color}
\usepackage{cancel}
\usepackage{wrapfig}
\usepackage{caption}
\usepackage{xfrac}
\usepackage{algorithm}
\usepackage{algorithmic}
\usepackage{mdframed}
\usepackage{hyperref}
\usepackage{cleveref}
\usepackage{todonotes}
\usepackage{tikz,pgfplots}
\usepackage{xcolor}
\hypersetup{
	colorlinks,
	linkcolor={red!40!gray},
	citecolor={blue!40!gray},
	urlcolor={blue!70!gray}
}
\def\Vol{{\mathrm{Vol}}}
\def\DPP{{\mathrm{DPP}}}

\def\poly{{\mathrm{poly}}}
\def\polylog{{\mathrm{polylog}}}

\newcommand{\MSE}[1] {{\mathrm{MSE}\big[#1\big]}}

\newcommand{\Er}{\mathrm{Er}}

\def\Rb{{\mathbf{R}}}

\def\xib{\boldsymbol\xi}

\def\xbt{\widetilde{\x}}

\def\Xt{\widetilde{X}}

\def\Rb{\mathbf R}
\def\Rbt{\mathbf{\widetilde{R}}}

\def\L{\mathbf{L}}

\def\Nc{\mathcal{N}}

\def\R{\mathbf R}

\def\K{\mathbf K}

\ifx\BlackBox\undefined
\newcommand{\BlackBox}{\rule{1.5ex}{1.5ex}}  
\fi
\DeclareMathOperator*{\argmin}{\mathop{\mathrm{argmin}}}

\def\x{\mathbf x}

\def\y{\mathbf y}

\def\ybt{\widetilde{\mathbf y}}

\def\yt{\widetilde{y}}

\def\w{\mathbf w}
\def\v{\mathbf v}

\def\wbh{\widehat{\mathbf w}}
\def\wbt{\widetilde{\mathbf w}}
\def\e{\mathbf e}
\def\zero{\mathbf 0}

\def\u{\mathbf u}

\def\X{\mathbf X}

\def\B{\mathbf B}

\def\A{\mathbf A}

\def\U{\mathbf U}

\def\Ubt{\widetilde{\mathbf U}}

\def\S{\mathbf S}

\def\St{\widetilde{S}}

\def\I{\mathbf I}

\def\A{\mathbf A}

\def\Xt{\widetilde{\mathbf{X}}}

\def\E{\mathbb E}

\def\R{\mathbb R} 
\def\tr{\mathrm{tr}}

\def\rank{\mathrm{rank}}

\let\origtop\top
\renewcommand\top{{\scriptscriptstyle{\origtop}}} 

\definecolor{silver}{cmyk}{0,0,0,0.3}
\definecolor{yellow}{cmyk}{0,0,0.9,0.0}
\definecolor{reddishyellow}{cmyk}{0,0.22,1.0,0.0}
\definecolor{black}{cmyk}{0,0,0.0,1.0}
\definecolor{darkYellow}{cmyk}{0.2,0.4,1.0,0}
\definecolor{darkSilver}{cmyk}{0,0,0,0.1}
\definecolor{grey}{cmyk}{0,0,0,0.5}
\definecolor{darkgreen}{cmyk}{0.6,0,0.8,0}

\newcommand{\Green}[1]{{\color{darkgreen}  {#1}}}
\newcommand{\Blue}[1]{\color{blue}{#1}\color{black}}
\newcommand{\Brown}[1]{{\color{brown}{#1}\color{black}}}
\newcommand{\white}[1]{{\textcolor{white}{#1}}}

\ifx\proof\undefined
\newenvironment{proof}{\par\noindent{\bf Proof\ }}{\hfill\BlackBox\\[2mm]}
\fi

\ifx\theorem\undefined
\newtheorem{theorem}{Theorem}
\fi

\ifx\example\undefined
\newtheorem{example}{Example}
\fi

\ifx\property\undefined
\newtheorem{property}{Property}
\fi

\ifx\lemma\undefined
\newtheorem{lemma}{Lemma}
\fi

\ifx\proposition\undefined
\newtheorem{proposition}{Proposition}
\fi

\ifx\remark\undefined
\newtheorem{remark}{Remark}
\fi

\ifx\corollary\undefined
\newtheorem{corollary}{Corollary}
\fi

\ifx\definition\undefined
\newtheorem{definition}{Definition}
\fi

\ifx\conjecture\undefined
\newtheorem{conjecture}[theorem]{Conjecture}
\fi

\ifx\axiom\undefined
\newtheorem{axiom}[theorem]{Axiom}
\fi

\ifx\claim\undefined
\newtheorem{claim}[theorem]{Claim}
\fi

\ifx\assumption\undefined
\newtheorem{assumption}{Assumption}
\fi

\title{
Determinantal Point Processes in\\ Randomized
Numerical Linear Algebra
}
\author{%
\textbf{Micha{\l } Derezi\'{n}ski} \\
Department of Statistics\\
University of California, Berkeley\\
\texttt{mderezin@berkeley.edu}
\and
\textbf{Michael W. Mahoney}\\
ICSI and Department of Statistics\\
University of California, Berkeley\\
\texttt{mmahoney@stat.berkeley.edu}
}
\date{}

\begin{document}
\maketitle

\begin{abstract}
Randomized Numerical Linear Algebra (RandNLA) uses
randomness to develop improved algorithms for matrix
problems that arise in scientific computing, data science,
machine learning, etc. Determinantal Point Processes (DPPs), a
seemingly unrelated topic in pure and applied mathematics, is a class
of stochastic point processes with probability distribution characterized by
sub-determinants of a kernel matrix.  Recent work has uncovered deep
and fruitful connections between DPPs and RandNLA which lead to new
guarantees and improved algorithms that are of interest to both areas.
We provide an overview of this exciting new line of research,
including brief introductions to RandNLA and DPPs, as well as
applications of DPPs to classical linear algebra tasks such as least
squares regression, low-rank approximation and the Nystr\"om method.
For example, random sampling with a DPP leads to new kinds of unbiased
estimators for least squares, enabling more refined statistical and
inferential understanding of these algorithms; a DPP is, in some
sense, an optimal randomized algorithm for the Nystr\"om method; and a
RandNLA technique called leverage score sampling can be derived as the
marginal distribution of a DPP.  We also discuss recent algorithmic
developments, illustrating that, while not quite as efficient as
standard RandNLA techniques, DPP-based algorithms are only
moderately more expensive. 
\end{abstract}

\section{Introduction}
\label{s:intro}

Randomized Numerical Linear Algebra (RandNLA), is an area which uses randomness, most notably random sampling and random projection methods, to develop improved algorithms for ubiquitous matrix problems.  
It began as a niche area in theoretical computer science about fifteen years ago~\cite{drineas2006multiplication,dkm_matrix2,dkm_matrix3}, and since then the area has exploded.
Matrix problems are central to much of applied mathematics, from traditional scientific computing and partial differential equations to statistics,  machine learning, and artificial intelligence.
Generalizations and variants of matrix problems are central to many other areas of mathematics, via more general transformations and algebraic structures, nonlinear optimization, infinite-dimensional operators, etc. 
Much of the work in RandNLA has been propelled by recent developments in machine learning, artificial intelligence, and large-scale data science, and RandNLA both draws upon and contributes back to both pure and applied mathematics. 

A seemingly different topic, but one which has a long history in pure and applied mathematics, is that of Determinantal Point Processes (DPPs).
A DPP is a stochastic point process, the probability distribution of which is characterized by sub-determinants of some matrix. 
Such processes were first studied to model the distribution of fermions at thermal equilibrium~\cite{dpp-physics}. 
In the context of random matrix theory, DPPs emerged as the eigenvalue distribution for standard random matrix ensembles~\cite{Bor11}, and they are of interest in other areas of mathematics such as graph theory, combinatorics and quantum mechanics~\cite{BO00,dpp-independence}.   
More recently, DPPs have also attracted significant attention within machine learning and statistics as a tractable probabilistic model that is able to capture a balance between quality and diversity within data sets and that admits efficient algorithms for sampling, marginalization, conditioning, etc.~\cite{dpp-ml}. 
This resulted in practical application of DPPs in experimental design
\cite{symmetric-polynomials}, recommendation systems \cite{k-dpp},
stochastic optimization \cite{dpp-minibatch} and more.

Until very recently, DPPs have had little if any presence within RandNLA.
However, recent work has uncovered deep connections between these two topics.
The purpose of this article is to provide an overview of RandNLA, with an emphasis on discussing and highlighting these connections with DPPs. 
In particular, we will show how random sampling with a DPP leads to new kinds of unbiased estimators for the classical RandNLA task of least squares regression, enabling a more refined statistical and inferential understanding of RandNLA algorithms. 
We will also demonstrate that a DPP is, in some sense, an optimal randomized method for low-rank approximation, another ubiquitous matrix problem. 
Finally, we also discuss how a standard RandNLA technique, called leverage score sampling, can be derived as the marginal distribution  of a DPP, as well as the algorithmic consequences this has for efficient DPP sampling.

We start (in Section~\ref{sxn:randnla}) with a brief review of a prototypical RandNLA algorithm, focusing on the ubiquitous least squares problem and highlighting key aspects that will put in context the recent work we will review. 
In particular, we discuss the trade-offs between standard sampling methods from RandNLA, including uniform sampling, norm-squared sampling, and leverage score sampling. 
Next (in Section~\ref{sxn:dpps}), we introduce the family of DPPs, highlighting some important sub-classes and the basic properties that make them appealing for RandNLA.
Then (in Section~\ref{sxn:applications}), we describe the fundamental connections between certain classes of DPPs and the classical RandNLA tasks of least squares regression and low-rank approximation, as well as the relationship between DPPs and the RandNLA method of leverage score sampling. 
Finally (in Section~\ref{sxn:algorithms}), we discuss the algorithmic aspects of both leverage scores and DPPs.
We conclude (in Section~\ref{sxn:conc}) by briefly mentioning several
other connections between DPPs and RandNLA, as well as a
recently introduced class of random matrices, called determinant preserving,
which has proven useful in this line of research.

\section{RandNLA: Randomized Numerical Linear Algebra}
\label{sxn:randnla}

In a typical RandNLA setting, we are given a large dataset in the form of a real-valued matrix, say $\X\in\R^{n\times d}$, and our goal is to compute quantities of interest quickly~\cite{Mah-mat-rev_BOOK,DM16_CACM}.
To do so, we efficiently down-size the matrix using a randomized algorithm, while approximately preserving its inherent structure, as measured by some metric. 
In doing so, we obtain a new matrix $\Xt$ (often called a \emph{sketch} of $\X$) which is either smaller or sparser than the original matrix.
Many applications of RandNLA follow this \emph{sketch-and-solve} paradigm: instead of performing a costly operation on $\X$,  we first construct $\Xt$ (the \emph{sketch}); we then perform the expensive operation (more cheaply) on the smaller $\Xt$ (the \emph{solve}); and we use the solution from $\Xt$ as a proxy for the solution we would have obtained from $\X$.
Here, cost often means computational time, but it can also refer to communication or storage space or even human work. 
The success of RandNLA methods has been proven in many domains, e.g., when the randomized least squares solvers such as Blendenpik \cite{blendenpik} or LSRN~\cite{MSM14_SISC} have outperformed the established high performance computing software LAPACK or other methods in parallel/distributed environments, respectively, or when RandNLA methods have been used in conjunction with traditional scientific computing solvers for low-rank approximation problems~\cite{HMT09_SIREV}.

Many different approaches have been established for randomly down-sizing data matrices $\X$ (see \cite{Mah-mat-rev_BOOK,DM16_CACM} for a detailed survey).
While some methods randomly zero out most of the entries of the matrix, most randomly keep only a small random subset of rows and/or columns.
In either case, however, the choice of randomness is crucial in preserving the structure of the data. 
For example, if the data matrix $\X$ contains a few dominant entries/rows/columns (e.g., as measured by their absolute value or norm or some other ``importance'' score), then we should make sure that our sketch is likely to retain the information they carry. 
This leads to data-dependent sampling distributions that will be the focus of our discussion. 
However, data-independent sketching techniques, which involve applying
a random transformation $\S$ to the matrix $\X$ (typically called a
``random rotation'' or a ``random projection," even if it is not precisely a rotation or projection in the linear algebraic sense), have also proven very successful \cite{Woodruff_sketching_NOW}. 
Among the most common examples of such random transformations are
i.i.d.~Gaussian matrices, fast Johnson-Lindenstraus transforms
\cite{ailon2009fast} and count sketches \cite{cw-sparse}, all of which provide different
trade-offs between efficiency and accuracy.
These ``data-oblivious random projections'' can be interpreted either in terms of the Johnson-Lindenstraus lemma \cite{johnson-lindenstrauss} or as a preconditioner for the ``data aware random sampling'' methods we discuss \cite{Mah-mat-rev_BOOK,DM16_CACM}.

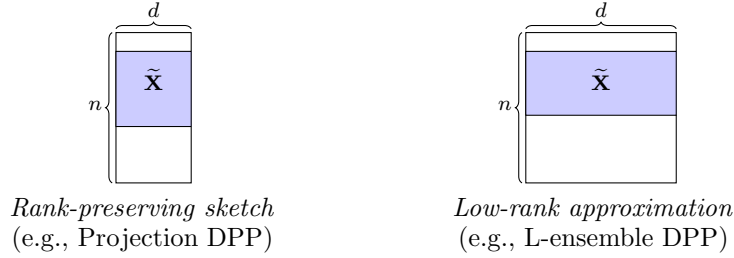
\begin{figure}
  \centering
\begin{tikzpicture}[scale=0.5]
  \draw (0,0) rectangle (2,4);
    \draw[fill=blue!20] (0,1.5) rectangle (2,3.5);
    \draw (1,2.75) node {\mbox{\footnotesize $\Xt$}};
    \draw [decorate,decoration={brace}] (-.1,0) -- (-.1,4);
    \draw (-.55,2.05) node {\mbox{\fontsize{8}{8}\selectfont $n$}}; 
    \draw [decorate,decoration={brace}] (0,4.1) -- (2,4.1);
    \draw (1,4.6) node {\mbox{\fontsize{8}{8}\selectfont $d$}};
    \draw (.7,-.7) node {\emph{Rank-preserving sketch}};
    \draw (.7,-1.5) node {(e.g., Projection DPP)};
  \end{tikzpicture}
\ifisarxiv \hspace{2cm}\else  \hfill\fi
  \begin{tikzpicture}[scale=0.5]
    \draw (0,0) rectangle (4,4);
    \draw[fill=blue!20] (0,1.8) rectangle (4,3.5);
    \draw (2,2.75) node {\mbox{\footnotesize $\Xt$}};
    \draw [decorate,decoration={brace}] (-.1,0) -- (-.1,4);
    \draw (-.55,2.05) node {\mbox{\fontsize{8}{8}\selectfont $n$}}; 
    \draw [decorate,decoration={brace}] (0,4.1) -- (4,4.1);
    \draw (2,4.6) node {\mbox{\fontsize{8}{8}\selectfont $d$}}; 
    \draw (1.8,-.7) node {\emph{Low-rank approximation}};
    \draw (1.8,-1.5) node {(e.g., L-ensemble DPP)};
  \end{tikzpicture}
\caption{Two RandNLA settings which are differentiated by whether the
  sketch (matrix $\Xt$) aims to preserve the rank of~$\X$ (left)
  or obtain a low-rank approximation (right). In Section
  \ref{sxn:applications} we associate each setting with a different DPP. }
\label{f:settings}
\end{figure}

Most RandNLA techniques can be divided into one of two settings, depending on the dimensionality or aspect ratio of $\X$, and on the desired size of the sketch (see Figure \ref{f:settings}):
\begin{enumerate}
\item \underline{Rank-preserving sketch}. When $\X$ is a tall
  full-rank matrix (i.e., $n\gg d$), then we can reduce the larger
  dimension while preserving the rank. 
\item \underline{Low-rank approximation}. When $\X$ has
comparably large dimensions (i.e., $n\sim d$), then the sketch
  typically has a much lower rank than $\X$.
\end{enumerate}
The classical application of rank-preserving sketches is least squares regression, where, given matrix $\X\in\R^{n\times d}$ and vector $\y\in\R^n$, we wish to find:
\begin{align*}
  \w^* = \argmin_\w L(\w),\ \text{ for }\ L(\w)=\|\X\w-\y\|^2.
\end{align*}
The least squares solution can be computed exactly, using the Moore-Penrose pseudoinverse, $\w^*=\X^\dagger\y$.
In a traditional RandNLA setup, in order to avoid solving the full problem, our goal is to use the sketch-and-solve paradigm to obtain
an $(\epsilon,\delta)$-approximation of $\w^*$, i.e., $\wbh$ such that:
\begin{align}
  L(\wbh)\leq (1+\epsilon)L(\w^*)\ \text{ with probability }\
  1-\delta.\label{eq:randnla-guarantee}
  \end{align}
Imposing statistical modeling assumptions on the vector $\y$ leads to different objectives, such as the mean squared error (MSE): 
\begin{align*}
  \MSE{\wbh} = \E\,\|\wbh-\wbt\|^2,\text{ given }\y=\X\wbt+\xib,
\end{align*}
where $\xib$ is a noise vector with a known distribution.
There has been work on statistical aspects of RandNLA methods \cite{MMY15,asymptotic_RandNLA_estimators_TR}, and these statistical objectives pose different challenges than the standard RandNLA guarantees (some of which can be addressed by DPPs, see Section \ref{sxn:applications}).

To illustrate the types of guarantees achieved by RandNLA methods on the least squares task, we will focus on row sampling, i.e., sketches consisting of a small random subset of the rows of $\X$, in the case that $n \gg d$.
Concretely, the considered meta-strategy is to draw random i.i.d.~row indices $j_1,...,j_k$ from $\{1,...,n\}$, with each index distributed according to $(p_1,...,p_n)$, and then solve the subproblem formed from those indices:
  \begin{align}
    \wbh = \argmin_{\w}\|\Xt\w-\ybt\|^2 = \Xt^\dagger\ybt, \label{eq:randnla-estimator}
  \end{align}
where
$\xbt_i^\top=\sqrt{\white{|}\!\!\!\smash{\text{\fontsize{9}{9}\selectfont$\tfrac1{kp_{j_i}}$}}}\,\x_{j_i}^\top$
  and
  $\yt_i=\sqrt{\white{|}\!\!\!\smash{\text{\fontsize{9}{9}\selectfont$\tfrac1{kp_{j_i}}$}}}\,y_{j_i}$,
  for $i=1,...,k$, denote the
$i^{th}$ row of $\Xt$ and entry of $\ybt$, respectively.
The rescaling is introduced to account for the biases caused by non-uniform sampling. 
We consider the following standard sampling distributions: 
\begin{enumerate}
  \item 
  \underline{Uniform}: 
  $p_i=1/n$, 
  for all $i$. 
  \item 
  \underline{Squared norms}: $p_i= \|\x_i\|^2 / \|\X\|_F^2$, 
 where $\|\cdot\|_F$ is the Frobenius (Hilbert-Schmidt) norm.
  \item 
  \underline{Leverage scores}: $p_i= l_i / d$, 
  where we let $l_i=\|\x_i\|^2_{(\X^\top\X)^{-1}}$ denote the $i^{th}$
  leverage score of $\X$ and $\|\v\|_\A=\sqrt{\v^\top\A\v}$ is the Mahalanobis norm.
    \end{enumerate}
Both squared norm and leverage score sampling are standard RandNLA techniques used in a variety of applications \cite{drineas2006sampling,drineas2006multiplication,cur-decomposition}. 
The following theorem (which, for convenience, we state with a failure probability of $\delta = 0.1$) puts together the results that allow us to compare each row sampling distribution in the context of least squares. 
\begin{theorem}\label{t:iid}
Estimator $\wbh$ constructed as in \eqref{eq:randnla-estimator} is an
$(\epsilon,0.1)$-approximation, as in \eqref{eq:randnla-guarantee}, if:
  \begin{enumerate}
    \item $k=\Omega(\mu d \log d + \mu d/\epsilon)$ for \underline{Uniform}, where $\mu$
      denotes the matrix coherence of $\X$.
    \item $k=\Omega(\kappa d\log d + \kappa d/\epsilon)$ for
      \underline{Squared norms}, where $\kappa$ is the condition number of $\X^\top\X$.
    \item $k=\Omega(d\log d + d/\epsilon)$ for \underline{Leverage scores}.
    \end{enumerate}
  \end{theorem}
Recall that (when considering least squares) we typically assume that $n\gg d$, so any of the three sample sizes $k$ may be much smaller than $n$. 
Thus, each sampling method offers a potentially useful guarantee for the number of rows needed to achieve a $(1+\epsilon)$-approximation.
However, in the case of both uniform and squared norm sampling, the sample size depends not only on the dimension $d$, but also on other data-dependent quantities. 
For uniform sampling, that quantity is \emph{matrix coherence} $\mu$, which measures the degree of non-uniformity among the data points in terms of their leverage scores: $\mu=\max_i\frac nd l_i$. 
For squared norm sampling, that quantity is the
\emph{condition number} $\kappa$, which is the ratio between the largest and the smallest eigenvalue of the $d\times d$ data covariance $\X^\top\X$. 
To address the dependence on the condition number, one can replace the standard Euclidean norm with a Mahalanobis norm that takes into account the data covariance, obtaining the leverage scores.

We now briefly discuss a key structural property, called the \emph{subspace embedding}, which is needed to show the guarantees of Theorem \ref{t:iid}. 
This important property---first introduced into RandNLA for data-aware random sampling by \cite{drineas2006sampling} and then for data-oblivious random projection by \cite{sarlos-sketching,DMMS07_FastL2_NM10}---is ubiquitous in the analysis of many RandNLA techniques. 
Remarkably, most DPP results do \emph{not} rely on subspace embedding techniques, which is an important differentiating factor for  this class of sampling distributions. 
\begin{definition}\label{d:subspace-embedding}
  A sketching matrix $\S$ is a $(1\pm\epsilon)$ subspace embedding for
  the column space of  $\X$ if:
  \begin{align*}
    (1-\epsilon)\|\X\v\|^2\leq    \|\S\X\v\|^2\leq
    (1+\epsilon)\|\X\v\|^2,\ \ \forall \v\in\R^d.
  \end{align*}
\end{definition}

\noindent
The matrix $\Xt$ used in \eqref{eq:randnla-estimator} for constructing
$\wbh$ can be written as $\Xt=\S\X$, by letting the $i^{th}$ row of
$\S$ be the scaled standard basis vector
$\sqrt{\white{|}\!\!\!\smash{\text{\fontsize{9}{9}\selectfont$\tfrac1{kp_{j_i}}$}}}\,\e_{j_i}$.
Relying on established measure concentration results for random
matrices (see, e.g., \cite{matrix-tail-bounds}), we can show that $\S$
is a subspace embedding (up to some failure probability) for each of
the i.i.d~sampling methods from Theorem \ref{t:iid}. 
However, only leverage score sampling (or random projections, which
precondition the input to have approximately uniform leverage scores
\cite{DMMS07_FastL2_NM10,DM16_CACM}) achieves this for $O(d\log d)$
samples, independent of any input-specific quantities such as the
coherence of condition number. 

The least squares task formulated in
\eqref{eq:randnla-guarantee}, as well as the subspace embedding
condition, require using rank-preserving 
sketches. However, these ideas can be naturally extended to the task
of low-rank approximation. In particular, as discussed in more 
detail in Section \ref{s:optimal}, the standard error metric for
low-rank sketches can be reduced to the least squares error defined on
a subspace with the same rank as the sketch~\cite{cur-decomposition}. Similarly, leverage score
sampling has been extended to adapt to the
low-rank setting. In Section \ref{s:marginals}, we discuss one of
these extensions, called \emph{ridge leverage scores}, and its
connections to DPPs. 

In the following sections, we show how non-i.i.d.~sampling via DPPs goes beyond the standard RandNLA analysis.
Among other things, this will permit us to obtain approximation
guarantees with fewer than $d\log d$ samples and without a failure
probability. 

\section{DPPs: Determinantal Point Processes}
\label{sxn:dpps}

\begin{figure}
    \centering
    \begin{tikzpicture}
      \draw [fill=purple!10,opacity=.5] (3.25,2.2) ellipse (4 and 2);
      \draw (3.25,3.95) node {SR measures};                              
      \draw [fill=blue!70,opacity=.5] (2,2) ellipse (2 and 1.5);
      \draw (2,3.3) node {DPPs};
      \draw [fill=blue!10,opacity=.5] (2,2) ellipse (1.47 and 1.05);
      \draw (2,2.5) node {L-ensembles};                        
      \draw [fill=red!50,opacity=.5] (5,2) ellipse (1.47 and 1.05);
      \draw (5,2.5) node {$k$-DPPs};
      \draw [thick,<-] (3.8,2) -- (3.9,3.15);
      \draw (4.75,3.3) node {{\small Projection DPPs}};
    \end{tikzpicture}
    \caption{A diagram illustrating different classes of determinantal distributions within a broader class of Strongly Rayleigh (SR) measures: DPPs (Definition~\ref{d:dpp}), L-ensembles (Definition~\ref{d:l-ensemble}), $k$-DPPs (Definition~\ref{d:k-dpp}) and Projection DPPs (Remark~\ref{r:projection}).}
    \label{f:dpp-diagram}
\end{figure}
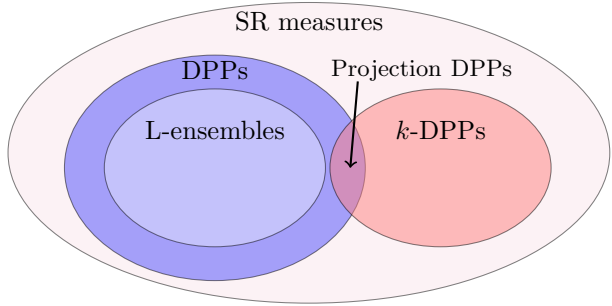

In this section, we define DPPs and related families of distributions (see Figure \ref{f:dpp-diagram} for a diagram), including some basic properties and intuitions. 
A detailed introduction to DPPs can be found in \cite{dpp-ml}. 
For a more general treatment that includes sampling from continuous domains, see \cite{dpp-independence}.

\begin{definition}[Determinantal Point Process]\label{d:dpp}
  Let $\K$ be an $n\times n$ positive semi-definite (p.s.d.) matrix such that $\zero\preceq\K\preceq\I$. 
  Point process $S\subseteq[n]$ is drawn according to $\DPP(\K)$, denoted as $S\sim\DPP(\K)$, if for any $T\subseteq[n]$,
  \begin{align*}
    \Pr\{T\subseteq S\} = \det(\K_{T,T}).
  \end{align*}
\end{definition}
Here, $\K_{T,T}$ denotes the $|T|\times |T|$ submatrix indexed by the set $T$.
Matrix $\K$ is called the \emph{marginal kernel} of $S$. 
If $\K$ is diagonal, then $\DPP(\K)$ corresponds to a series of $n$ independent biased coin-flips deciding whether to include each index $i$ into the set $S$. 
A more interesting distribution is obtained for a general $\K$, in which case the inclusion events are no longer independent.
Some of the key properties that make DPPs useful as a mathematical framework are: 
\begin{enumerate}
  \item \underline{Negative correlation}: if $i\neq j$ and
    $\K_{ij}\neq 0$, then $\Pr(i\in S \mid j\in S) < \Pr(i \in S)$.
  \item \underline{Cardinality}: while the size $|S|$ is in
    general random, its expectation equals $\tr(\K)$ and
    the variance also has a simple expression.
  \item \underline{Restriction}: the intersection $\St=S\cap T$ is distributed as
    $\DPP(\K_{T,T})$ (after relabeling).
  \item \underline{Complement}: 
    the complement set $\St=[n]\backslash S$ is distributed as $\DPP(\I-\K)$.
\end{enumerate}
In the context of linear algebra, a slightly more restrictive
definition of DPPs has proven useful.
\begin{definition}[L-ensemble]\label{d:l-ensemble}
Let $\L$ be an $n\times n$ p.s.d. matrix. Point process $S\subseteq[n]$ is drawn
according to $\DPP_L(\L)$ and called an L-ensemble if
\[\Pr\{S\} = \frac{\det(\L_{S,S})}{\det(\I+\L)}.\]
\end{definition}
It can be shown that any L-ensemble is a DPP by setting
$\K=\L(\I+\L)^{-1}$ (but not vice versa).
Unlike Definition \ref{d:dpp}, this definition explicitly gives the probabilities of
individual sets.
These probabilities sum to one as a consequence of a classical
determinantal identity (see Theorem 2.1 in \cite{dpp-ml}).
Furthermore, this direct formulation 
provides a natural geometric interpretation in the context of row
sampling for RandNLA. Suppose that we let $\L=\X\X^\top$ for some
$n\times d$ matrix $\X$. Then, the probability of sampling subset $S$
according to $\DPP_L(\L)$ satisfies:
\begin{align*}
\Pr\{S\} \propto \Vol^2\big(\{\x_i:i\in S\}\big)  .  
\end{align*}
Namely, this sampling probability is proportional to the squared $|S|$-dimensional volume of the parallelepiped spanned by the rows of $\X$ indexed by $S$. 
This immediately implies that the size of $S$ will never exceed the rank of $\X$ (which is bounded by $d$). 
Furthermore, such a distribution ensures that the set of sampled rows will be non-degenerate: no row can be obtained as a linear combination of the others. 
Intuitively, this property is desirable for RandNLA sampling as it avoids redundancies. 
Also, all else being equal, rows with larger norms are generally preferred as they contribute more to the volume. 

While the subset size of a DPP is in most cases a random variable, it is easy to constrain the cardinality to some fixed value $k$. 
The resulting distribution is not a DPP, in the sense of Definition~\ref{d:dpp}, but it retains many useful properties of proper DPPs.

\begin{definition}[Cardinality constrained DPP]\label{d:k-dpp}
We will use $k$-$\DPP_L(\L)$ to denote a
distribution obtained by constraining $\DPP_L(\L)$ to only
subsets of size $|S|=k$.
\end{definition}

\noindent
While a $k$-DPP is not, in general, a DPP in the sense of Definition \ref{d:dpp}, both families belong to a broader class of negatively correlated distributions called Strongly Rayleigh (SR) measures, which were introduced by \cite{Borcea09}. 
See Figure~\ref{f:dpp-diagram}.

At the intersection of DPPs and $k$-DPPs lies a family of distributions called Projection DPPs.
This family is of particular importance to RandNLA.
\begin{remark}[Projection DPP]\label{r:projection}
Point process $S\sim k$-$\DPP_L(\L)$ satisfies Definition \ref{d:dpp} iff
 $k=\rank(\L)$, in which case, we call it a Projection DPP.
\end{remark}

\noindent
Alternatively, a Projection DPP can also be defined as a $\DPP(\K)$ such that the marginal kernel $\K$ is an orthogonal projection matrix, hence the name. 

We chose to introduce Projection DPPs via the connection to L-ensembles to highlight once again the geometric interpretation. 
In this viewpoint, letting $\L=\X\X^\top$ and $k=\rank(\X)$, a Projection DPP associated with the L-ensemble $\L$ has marginal kernel $\K=\X \X^\dagger$, which is a $k$-dimensional projection onto the span of the columns of $\X$. 
Furthermore, the probability of a row subset of $\X$ under $S\sim\DPP(\X\X^\dagger)$ is proportional to the squared $k$-dimensional volume spanned by it. 
This implies that the rows $\{\x_i:i\in S\}$ sampled from this distribution will with probability 1 capture all directions of the ambient space that are present in the matrix $\X$.
This is an important property in the context of rank-preserving
sketches, as we will see next.


\section{DPPs in RandNLA}
\label{sxn:applications}

In this section, we demonstrate the fundamental connections between DPPs and standard RandNLA tasks, as well as the new kinds of RandNLA guarantees that can be achieved via these connections. 
Our discussion focuses on two types of DPP-based sketches (that were illustrated in Figure \ref{f:settings}):
\begin{enumerate}
\item Projection DPPs as a \emph{rank-preserving sketch};
\item L-ensemble DPPs as a \emph{low-rank approximation}.
\end{enumerate}
 These sketches can be efficiently constructed using DPP sampling
 algorithms which we discuss later (in Section \ref{sxn:algorithms}).
We also discuss the close relationship between DPPs and the RandNLA
method of leverage score sampling, shedding light on why these two
different randomized techniques have proven effective in RandNLA. For
the summary, see Table \ref{tab:comp}. 

\begin{table*}
  \centering
  \small
  \begin{tabular}{rl|rl|rl}
    &&\multicolumn{2}{l|}{\underline{\textit{Rank-preserving sketch}}}
    & \multicolumn{2}{l}{\underline{\textit{Low-rank approximation}}}\\[2mm]
&& Projection DPP & $S\!\sim\! d\text{-}\DPP_L(\X\X^\top)$\!\!& L-ensemble &$S\!\sim\!\DPP_L(\frac1\lambda\X\X^\top)$\\
  \hline\hline
subset size &$\E\,|S|\hfill=\!\!$& dimension&$d$ &effective dim.~&$\tr(\X\X^\top(\X\X^\top\!+\lambda\I)^{-1})$\\
  marginal &$\Pr\{i\in S\}\hfill=\!\!$&leverage score&
$\x_i^\top(\X^\top\X)^{-1}\x_i$
& ridge lev.~score &$\x_i^\top(\X^\top\X+\lambda\I)^{-1}\x_i$\\
~\!\!\!\!expectation &$\E\,\X_S^\dagger\y_S\hfill=\!\!$
& least squares &$\underset{\w}{\argmin}\,\|\X\w-\y\|^2$\!\!
       & ridge regression
& $\underset{\w}{\argmin}\,\|\X\w-\y\|^2\!+\lambda\|\w\|^2$\!\!\!\!
\end{tabular}
\caption{Key properties of the DPPs discussed in Section
  \ref{sxn:applications}, as they relate to: RandNLA tasks of least
  squares and ridge regression; RandNLA methods of leverage score
  sampling and ridge leverage score sampling.}
\label{tab:comp}
\end{table*}

\subsection{Unbiased estimators}
\label{s:expectations}

We now define the least squares estimators that naturally arise from
row sampling with Projection DPPs and L-ensembles. The definitions are motivated
by the fact that the estimators are unbiased, relative to the
solutions of the full least squares problems.  
Importantly, this property is \emph{not} shared by
i.i.d.~row sampling methods used in RandNLA.

We start with the rank-preserving setting, i.e., given a tall full-rank $n\times d$ matrix $\X$ and a vector $\y\in\R^n$, where $n\gg d$, we wish to approximate the least squares solution $\w^*=\argmin_\w\|\X\w-\y\|^2$. 
To capture all of the directions present in the data and to obtain a meaningful estimate of $\w^*$, we must sample at least $d$ rows from $\X$. 
We achieve this by sampling from a Projection DPP defined as $S\sim d$-$\DPP_L(\X\X^\top)$ (see Remark \ref{r:projection}). 
The linear system $(\X_S,\y_S)$, corresponding to the rows of $\X$ indexed by $S$, has exactly one solution because sets selected by a Projection DPP are always rank-preserving: $\wbh = \X_S^{-1}\y_S$.
Moreover, as shown by \cite{unbiased-estimates-journal}, the obtained random vector is an unbiased estimator of the least squares solution~$\w^*$.
\begin{theorem}\label{t:unbiased-ls}
  If $S\sim  d$-$\DPP_L(\X\X^\top)$, then
\begin{align*}
  \E\,\X_S^{-1}\y_S\ =\ \argmin_\w\|\X\w-\y\|^2 = \w^*.
\end{align*}
\end{theorem}

\noindent
This seemingly simple identity relies on the negative correlations
between the samples in a DPP, and thus \emph{cannot} hold for any
i.i.d.~row sampling method.

Theorem \ref{t:unbiased-ls} has an analogue in the context of low-rank 
approximation, where both the dimensions of $\X$ are comparably large
(i.e., $n\sim d$), and so the desired sample size is typically much
smaller than $d$. When the 
selected subproblem $(\X_S,\y_S)$ has fewer 
than $d$ rows (i.e., it is under-determined) then it has multiple
exact solutions. A standard way to address this is picking the 
solution with smallest Euclidean norm, defined via the Moore-Penrose
inverse: $\wbh = \X_S^\dagger\y_S$. 
To sample the under-determined subproblem, we use a scaled L-ensemble
DPP with the expected sample size controlled by a parameter
$\lambda>0$ \cite{surrogate-design}. 
\begin{theorem}\label{t:unbiased}
If $S\sim \DPP_L(\frac1\lambda\X\X^\top)$, then
\begin{align*}
  \E\, \X_S^\dagger\y_S \
  &=\
\argmin_\w \|\X\w-\y\|^2 + \lambda\|\w\|^2.
\end{align*}
\end{theorem}

\noindent
Thus, the minimum norm solution of the under-determined subproblem is
an unbiased estimator of the Tikhonov-regularized least squares problem,
i.e., ridge regression. Ridge regression is a natural extension of the
standard least squares task, particularly useful when $n\sim d$ or $n \ll d$.

Theorem \ref{t:unbiased} illustrates the
\emph{implicit regularization} effect that occurs when choosing one out of
many exact solutions to a subsampled least squares task (see also
Section~\ref{sxn:conc}). 
Increasing the regularization $\lambda\|\w\|^2$ in ridge regression is
interpreted as using fewer degrees of freedom, which
aligns with the effect that $\lambda$ has on the distribution
$S\sim\DPP_L(\frac1\lambda\X\X^\top)$: larger $\lambda$ means that
smaller subsets $S$ are more likely. In fact, the expected subset size of the
L-ensemble captures the notion of \emph{effective dimension}
(a.k.a.~effective degrees of freedom) in the same way as it is
commonly done for ridge regression in statistics
\cite{ridge-leverage-scores}:  
  \begin{align*}
d_\lambda &:=
             \tr\big(\X\X^\top\! (\X\X^\top\!+\lambda\I)^{-1}\big) = \E\,|S|.
  \end{align*}

\subsection{Exact error analysis}
The error analysis for DPP sampling  differs significantly from the standard RandNLA techniques discussed in Section \ref{sxn:randnla}.
In particular, approximation guarantees are formulated in terms of the expected error, without relying on measure concentration results. 
This means that we avoid failure probabilities such as the one present in Theorem \ref{t:iid}, and the analysis is often much more precise, sometimes even exact. 
Furthermore, because of the non-i.i.d.~nature of DPPs, these guarantees can be achieved with smaller sample sizes than for RandNLA sampling methods. 
Of course, this comes with computational trade-offs, which we discuss in Section \ref{sxn:algorithms}.

We illustrate these differences in the context of \emph{rank-preserving sketches} for least squares regression. 
Consider the estimator $\wbh=\X_S^{-1}\y_S$ from
Theorem~\ref{t:unbiased-ls}, where the subset $S$ is sampled via the
Projection DPP, i.e.,  $S\sim d$-$\DPP_L(\X\X^\top)$.
Recall that the sample size here is only $d$, which is less than
$d\log d$ needed by i.i.d.~sampling methods such as leverage scores  
(or random projection methods
\cite{DMMS07_FastL2_NM10,Woodruff_sketching_NOW}). 
Nevertheless, this estimator achieves an approximation guarantee in
terms of the expected loss.  Moreover, under minimal assumptions on
$\X$, the expectation can be given by a closed form expression \cite{unbiased-estimates}. 
\begin{theorem}\label{t:loss}
Assume that the rows of $\X$ are in general position, i.e., every set
of $d$ rows is non-degenerate. If $S\sim d$-$\DPP_L(\X\X^\top)$, then
\begin{align*}
\E\, L(\X_S^{-1}\y_S)  = (d+1)\,L(\w^*),
\end{align*}
and factor $d+1$ cannot in general be improved.
\end{theorem}

\noindent
This exact error analysis is particularly useful in statistical modeling, where under additional assumptions about the vector $\y$, we wish to estimate accurately the generalization error of our estimator. 
Specifically, consider the following noisy linear model of the vector $\y$:
\begin{align}
  \y = \X\wbt + \xib,\quad \text{where}\quad \xib\sim \Nc(\zero,\sigma^2\I).\label{eq:noise}
\end{align}
Here, $\wbt$ is the underlying linear transformation that we wish to recover, and mean squared error in this context is defined as $\E\,\|\wbh-\wbt\|^2$, where the expectation is taken over both the sampling and the noise. 
The Projection DPP estimator again enjoys an exact expression for the
mean squared error \cite{avron-boutsidis13,unbiased-estimates-journal}.
\begin{theorem}\label{t:mse}
Assume that the rows of $\X$ are in general position, i.e., every set of $d$ rows is non-degenerate, and consider $\y$ as in \eqref{eq:noise}. 
If $S\!\sim\!d$-$\DPP_L(\X\X^\top)$, then
\begin{align*}
\E\,\|\X_S^{-1}\y_S - \wbt\|^2  = (n-d+1)\,\E\,\|\w^*\!-\wbt\|^2.
\end{align*}
\end{theorem}
A number of extensions to Theorems~\ref{t:loss}~and~\ref{t:mse}
have been proposed, covering larger sample sizes \cite{leveraged-volume-sampling,chen2017condition} as well as different
statistical models \cite{regularized-volume-sampling,surrogate-design}.

\subsection{\ifisarxiv\else\!\!\!\fi Optimal approximation guarantees}
\label{s:optimal}
As we have seen above, the non-i.i.d.~nature of DPP sampling can lead
to improved approximation guarantees, compared to standard RandNLA
methods, when we wish to minimize the size of the down-sampled problem.  
We next discuss this in the \emph{low-rank approximation} setting,
i.e., when $n\sim d$. Here, cardinality constrained L-ensembles are known to achieve \emph{optimal} $(1+\epsilon)$-approximation guarantees.

In Section~\ref{s:expectations}, we used low-rank
sketches to construct unbiased estimates for regularized least squares,
 given matrix $\X$ and a vector $\y$. However, even
without introducing $\y$, a natural low-rank approximation
objective for sketching $\X$ can be defined via a reduction to least
squares~\cite{cur-decomposition}. Namely, we can measure the
error in reconstructing the $i^{th}$ row of $\X$ by finding the best
fit among all linear combinations of the rows of the sketch
$\X_S$. Repeating this over all rows of $\X$, we get: 
\begin{align*}
  \Er(S)
  &= \sum_{i=1}^n
\overbrace{\min_{\w}\|\X_S^\top\w-\x_i\|^2}^{\text{least squares}}
=\big\|\X\X_S^\dagger\X_S-\X\big\|_F^2,
\end{align*}
where $\X_S^\dagger\X_S$ is the projection onto the span of
$\{\x_i:i\in S\}$. If the size of $S$ is equal to some target rank
$r$, then $\Er(S)$ is at least as large as the error of the best
rank $r$ approximation of $\X$, denoted $\X_{(r)}$ (obtained by
projecting onto the top $r$ right-singular vectors of~$\X$).
However, \cite{pca-volume-sampling,more-efficient-volume-sampling}
showed that using a cardinality constrained L-ensemble with
$k=r+r/\epsilon-1$ rows suffices to get within a
$1+\epsilon$ factor of the best rank $r$ approximation error.

\begin{theorem}\label{t:low-rank}
If $S\sim
k\text{-}\DPP_L(\X\X^\top)$, where the sample
size satisfies $k\geq r + r/\epsilon-1$, then
\begin{align*}
\E\, \Er(S)  \leq (1+\epsilon)\,\|\X_{(r)}-\X\|_F^2,
\end{align*}
and the size $r+r/\epsilon-1$ is worst-case optimal.
\end{theorem}

\noindent
The task of finding the subset $S$ that minimizes $\Er(S)$ is sometimes known as the Column Subset
Selection Problem \cite{BMD09_CSSP_SODA} (with $\X$ replaced by $\X^\top$).  
Some refinements to the above bounds, relying on additional information
about the spectrum of $\X$, have been provided by \cite{nystrom-multiple-descent}.
Similar $1+\epsilon$ guarantees are achievable with RandNLA sampling techniques such
as leverage scores \cite{cur-decomposition}.
However, those require sample sizes $k$ of at least $r\log r$, they
contain a failure probability, and they suffer from additional
constant factors due to less exact~analysis.

\paragraph{Nystr\"om method.}
The task of low-rank approximation is often formulated in the context of symmetric positive semi-definite (p.s.d.) matrices. 
Let $\L$ be an $n\times n$ p.s.d.~matrix. 
We briefly discuss how DPPs can be applied in this setting via the Nystr\"om method, which constructs a rank $k$ approximation of $\L$ by using the eigendecomposition of a small $k\times k$ submatrix $\L_{S,S}$ for some index subset $S$. 
\begin{definition}
 We define the Nystr\"om approximation of $\L$ based on a subset $S$
 as the $n \times n$ matrix $\widetilde\L(S)=\L_{\cdot,S}\L_{S,S}^\dagger\L_{S,\cdot}$.
\end{definition}

\noindent
Originally developed in the context of obtaining numerical solutions to integral equations \cite{nystrom1930}, this method has found applications in a number of areas such as machine learning \cite{Williams01Nystrom,revisiting-nystrom}, Gaussian Process regression \cite{sparse-variational-gp} and Indenpendent Component Analysis \cite{Bach2003}. 
The following result \cite{belabbas-wolfe09} follows similarly as Theorem \ref{t:low-rank} and also provides the optimal sample size. 
We use $\|\cdot\|_*$ to denote the nuclear (trace) norm and $\L_{(r)}$ as the best rank $r$ approximation.
\begin{theorem}\label{t:nystrom-fixed}
If $S\sim
k\text{-}\DPP_L(\L)$, where the sample
size satisfies $k\geq r+r/\epsilon-1$, then
  \begin{align*}
    \E\,\|\L-\widetilde\L(S)\|_* \leq
    (1+\epsilon)\, \|\L-\L_{(r)}\|_*,
  \end{align*}
  and the size $r+r/\epsilon-1$ is worst-case optimal.
\end{theorem}

\subsection{\ifisarxiv\else\!\!\!\!\fi Connections to RandNLA methods}
\label{s:marginals}

The natural applicability of DPPs in the RandNLA tasks of least squares regression and low-rank approximation discussed above raises the question of how DPPs relate to traditional RandNLA sampling methods used for this task. 
As discussed in Section \ref{sxn:randnla}, one of the main RandNLA
techniques for constructing rank-preserving sketches (i.e., relative
to a tall matrix $\X$ with $n\gg d$) is
i.i.d.~leverage score sampling.
(From this perspective, random projections can be seen as preprocessing or preconditioning the input so that leverage scores are approximately uniform, thereby enabling uniform sampling---in the randomly-transformed space---to be successfully used.)  
Even though leverage score sampling was developed independently of
DPPs, this method can be viewed as an i.i.d.~counterpart of the
Projection DPP from Theorem \ref{t:unbiased-ls}. 
\begin{theorem}
Let $\X$ be an $n\times d$ full rank matrix. For $S\sim
d$-$\DPP_L(\X\X^\top)$, the marginal probability of $i\in[n]$ 
is equal to the $i^{th}$ leverage score of $\X$:
\begin{align*}
  \Pr\{i\in S\}=\x_i^\top(\X^\top\X)^{-1}\x_i.
\end{align*}
\end{theorem}

\noindent
This fact can be easily derived from the DPP properties discussed in
Section \ref{sxn:dpps}. Recall that the marginal kernel of the Projection
DPP is the projection matrix
\begin{align*}
  \K=\X\X^\dagger=\X(\X^\top\X)^{-1}\X^\top.
\end{align*}
The marginal probabilities of this distribution lie on the diagonal of
the marginal kernel, which also contains the leverage scores of $\X$.

Thus, leverage score sampling can be obtained as a distribution constructed from the marginals of the Projection DPP. 
Naturally, when going from i.i.d.~to non-i.i.d.~sampling, we lose all the negative correlations between the points in a DPP sample, and therefore the expectation formulas and inequalities from the preceeding sections no longer hold for leverage score sampling. 
Furthermore, recall that to achieve
a rank-preserving sketch (e.g., for least squares) with leverage score sampling for a
full rank matrix $\X$ we require at least $d\log d$ rows (Theorem \ref{t:iid}), whereas a
Projection DPP generates only $d$ samples and also provides
a rank-preserving sketch (Theorem
\ref{t:loss}). This shows that losing the negative correlations costs us a
factor of $\log d$ in the sample size.

Another connection between leverage scores and Projection DPPs emerges in the reverse direction, i.e., going from i.i.d.~to non-i.i.d.~samples. 
Namely, a leverage score sample of size at least $2d\log
d$ contains a Projection DPP  with probability at least $1/2$  \cite{minimax-experimental-design}.  
\begin{theorem}\label{t:dlogd}
Let $j_1,j_2,...$ be a sequence of i.i.d.~leverage score
samples from matrix $\X$. There is a random set $T\subseteq
\{1,2,...\}$ of size $d$ s.t. $\max\{i\in T\}\leq 2d\log d$ with
probability at least $1/2$, and:
\begin{align*}
  \{j_i: i\in T\}\sim d\textnormal{-}\DPP_L(\X\X^\top).
  \end{align*}
\end{theorem}

\noindent
Many extensions of leverage scores have been proposed for use in the
low-rank approximation setting \cite{revisiting-nystrom} (i.e., when
$n\sim d$).  
Arguably the most popular one is called \emph{ridge leverage scores}
\cite{ridge-leverage-scores}. 
Ridge leverage scores can be recovered as the marginals of an
L-ensemble. 
\begin{theorem}
For $S\!\sim\!\DPP_L(\frac1\lambda\X\X^\top)$, the marginal probability of index $i\in[n]$ is equal to the $i^{th}$ $\lambda$-ridge leverage score of $\X$:
\begin{align*}
  \Pr\{i\in S\}=\x_i^\top(\X^\top\X+\lambda\I)^{-1}\x_i.
\end{align*}
\end{theorem}
The typical sample size required for low-rank approximation with ridge leverage scores is at least $d_\lambda\log d_\lambda$, where $d_\lambda$ is the ridge effective dimension and also the expected size of the L-ensemble. 
Once again, the logarithmic factor appears as a trade-off coming from i.i.d.~sampling.

A reverse connection analogous to Theorem \ref{t:dlogd}, i.e., going
from i.i.d.~to non-i.i.d.~sampling, can also be obtainined for ridge
leverage scores \cite{dpp-intermediate}, although only a weaker
version, with $O(d_\lambda^2)$ instead of $O(d_\lambda\log
d_\lambda)$, is currently known in this setting. 

\begin{theorem}\label{t:dsquared}
Let $j_1,j_2,...$ be a sequence of i.i.d.~$\lambda$-ridge leverage score
samples from matrix $\X$. There is a random set $T\subseteq
\{1,2,...\}$ such that $\max\{i\in T\}\leq 2d_\lambda^2$ with
probability at least $1/2$, and:
\begin{align*}
  \{j_i: i\in T\}\sim \DPP_L(\tfrac1\lambda\X\X^\top).
  \end{align*}
\end{theorem}

\section{Sampling algorithms}
\label{sxn:algorithms}

One of the key considerations in RandNLA is computational efficiency of constructing random sketches. 
For example, the i.i.d.~leverage score sampling sketch defined in Section \ref{sxn:randnla} requires pre-computing all of the leverage scores.
If done na\"ively, this costs as much as performing the singular value decomposition (SVD) of the data.
However, by employing fast RandNLA projection methods, one obtains efficient near-linear time complexity approximation algorithms for leverage score sampling \cite{fast-leverage-scores}. 

In the case of DPPs, the challenge may seem even more daunting, since the na\"ive algorithm has exponential time complexity relative to the data size. 
However, the connections between leverage scores and DPPs (summarized in Table \ref{tab:comp}) have recently played a crucial role in the algorithmic improvements for DPP sampling.
In particular, recent advances in DPP sampling have resulted in several algorithmic techniques which are faster than SVD, and in some regimes even approach the time complexity of fast leverage score sampling algorithms. 
See Table \ref{tab:algs} for an overview.

\subsection{Leverage scores: Approximation}
\label{s:fast-leverage}

We start by discussing fast sketching methods for approximating
leverage scores more rapidly than by na\"{\i}vely computing them via
the SVD or a QR decomposition.  
This is a good illustration of RandNLA algorithmic
techniques, and it is also relevant in our later discussion of DPP sampling.

For simplicity, we focus on constructing leverage scores for
rank-preserving sketches (i.e., for a tall $n\times d$ matrix $\X$), but similar ideas apply to the low-rank approximation setup~\cite{fast-leverage-scores}.
Recall that the $i^{th}$ leverage score of $\X$ is given by
$l_i=\x_i^\top(\X^\top\X)^{-1}\x_i$, which can be expressed as the
squared norm of the $i^{th}$ row of the matrix $\X(\X^\top\X)^{-\frac12}$.  
Assuming that $n\gg d$, the primary computational cost of obtaining
this matrix involves two expensive matrix multiplications: first,
computing $\Rb=\X^\top\X$ (or a similarly expensive operation such as
a QR decomposition or the SVD); and second, computing $\X\Rb^{-\frac12}$.
While each of these steps costs $O(nd^2)$ arithmetic operations,
\cite{fast-leverage-scores} showed that both of them can be
approximated using efficient randomized sketching techniques.  

The first step, which involves producing a matrix
$\Rbt=\Xt^\top\Xt\approx\Rb$, requires using 
a rank-preserving sketch $\Xt$, since we must ensure the invertibility
of $\Rbt$. This can be achieved, e.g., using the Subsampled Randomized
Hadamard Transform sketch \cite{ailon2009fast}, which is a random sketching
matrix $\S$ that, with high probability, satisfies the subspace
embedding property (Definition \ref{d:subspace-embedding}) on matrix $\X$.
Crucially, the matrix multiplication $\Xt=\S\X$ can be
performed in time $O(nd\log d)$. The resulting sketch $\Xt$ has
$O(d\log n)$ rows, so computing $\Rbt=\Xt^\top\Xt$ takes $O(d^3\log n)$~time.

The second step, i.e., sketching the matrix product
$\X\Rbt^{-\frac12}$, involves constructing a low-rank approximation that must preserve
all $n$ row-norms of the matrix. This property is satisfied by
sketching matrices known as Johnson-Lindenstrauss Transforms
\cite{johnson-lindenstrauss,achlioptas2003}. A simple i.i.d.~Gaussian matrix $\S$ with
$r=O(\log n)$ columns in $\R^n$ suffices here. If we compute the
$n\times r$ matrix $\X\Rbt^{-\frac12}\S$, then the row-norms of this
matrix with high probability provide good approximations for all of the
leverage scores of $\X$. 
A good approximation here means a value, say, $\tilde l_i\in[\frac12l_i,\frac32l_i]$. 
The cost of this step is $O(nd\log n)$. 
The resulting overall procedure returns leverage score approximations much faster than the na\"ive $O(nd^2)$ algorithm.

A number of refinements have been proposed
\cite{cw-sparse,NN13,iterative-row-sampling} for approximating leverage scores,
and similar approaches have also been developed for ridge leverage
scores \cite{ridge-leverage-scores,NIPS2018_7810}, which are used for
low-rank approximation. In this case, we often consider a kernelized 
version of the problem, where instead of an $n\times d$ matrix $\X$,
we are given an $n\times n$ kernel matrix $\K=\X\X^\top$ (which is
also particularly relevant in the context of DPPs). Here,
$\lambda$-ridge leverage scores can be defined as the diagonal entries of the
matrix $\K(\lambda\I+\K)^{-1}$ and can be approximately computed in
time $O(n d_\lambda^2\,\polylog(n))$. When $d_\lambda\ll n$, this is much less than the na\"ive
cost of $O(n^3)$.  

\begin{table*}
  \centering
  \small
  \begin{tabular}{rl|cc|cc}
    &&\multicolumn{2}{l|}{\underline{\textit{Rank-preserving sketch}}}&
\multicolumn{2}{l}{\underline{\textit{Low-rank approximation}}}\\[2mm]
 &   & \multicolumn{2}{l|}{\textit{Input:} $n\times d$ data matrix $\X$,
      $n\gg d$}
    & \multicolumn{2}{l}{\textit{Input:} $n\times n$ kernel matrix
      $\L$}\\[-.5mm]
&&\multicolumn{2}{l|}{\textit{Output:} Sample of size $k=O(d)$}
& \multicolumn{2}{l}{\textit{Output:} Sample of size $k\ll n$}
    \\[1.5mm]
&& First sample & Subsequent samples & First sample & Subsequent samples\\
    \hline\hline
Leverage scores: &exact & $nd^2$&$d$ & $n^3$ & $k$ \\
    &approximate & $nd+d^3$&$d$ & $nk^2$ & $k$ \\
    \hline
DPPs: &exact & $nd^2$ & $d^3$& $n^3$ & $nk + k^3$\\
&intermediate & $nd+\poly(d)$&$\poly(d)$& $n\cdot\poly(k)$ & $\poly(k)$\\
&Monte Carlo & $n\cdot\poly(d)$ & $n\cdot\poly(d)$& $n\cdot\poly(k)$ & $n\cdot\poly(k)$
  \end{tabular}
\caption{%
  Comparison of sampling cost for DPP algorithms, alongside the
  cost of exact and approximate leverage score sampling, given either
  a tall data matrix $\X$ or a square p.s.d.~kernel $\L$. Most methods
  can also be extended to the wide data matrix $\X$ setting.
  We assume that an L-ensemble kernel $\L$ is used
  for the DPPs (if given a data matrix $\X$, we use $\L=\X\X^\top$).
  We allow either a $k$-DPP or an L-ensemble with expected size $k$, however, in some cases,
  there are minor differences in the time complexities (in which case we give the
  better of the two). For simplicity, we omit the log terms in these expressions. }
\label{tab:algs}
\end{table*}

\subsection{DPPs: Exact sampling}
\label{s:mixture}
In this and subsequent sections, we discuss several algorithmic
techniques for sampling from DPPs and $k$-DPPs. We focus on the
general parameterization of a DPP via an $n\times n$ kernel matrix
(either the correlation kernel $\K$ or the L-ensemble kernel $\L$), but
we also discuss how these techniques can be applied to sampling
from DPPs defined on a tall $n\times d$ matrix~$\X$, which we used
in Section \ref{sxn:applications}. 

We start with an important result of
\cite{dpp-independence}, which shows that any DPP can be decomposed into a
mixture of Projection DPPs. 
\begin{theorem}\label{t:mixture}
Consider the eigendecomposition $\K=\sum_i\lambda_i
\u_i\u_i^\top$, where $\lambda_i\in[0,1]$ for all $i$ and let
$s_i\sim\mathrm{Bernoulli}(\lambda_i)$. Then the mixture distribution
$\DPP(\sum_is_i\u_i\u_i^\top)$ is identical to $\DPP(\K)$.
\end{theorem}
This result broadened the popularity of DPPs in the computer science community because of its algorithmic implications.
However, to complete it into a proper algorithm, we must still show how to sample from a Projection DPP. 
Let $\U$ denote the $n\times k$ matrix whose columns are the eigenvectors $\u_i$ that are randomly selected in Theorem \ref{t:mixture}. 
Recall from Section \ref{sxn:dpps} that if $S\sim \DPP(\U\U^\top)$, then the marginal probability of $i\in S$ is given by
\begin{align}
  \Pr\{i\in S\} = \big[\U\U^\top\big]_{ii} = \|\v_i\|^2,\label{eq:proj-marginals}
\end{align}
where $\v_i^\top$ denotes the $i^{th}$ row of the matrix $\U$.
(Note that since $\U$ is an orthogonal matrix, its squared row-norms are also its leverage scores.) 
This gives us an easy way to sample the first point to be included in $S$.
The key idea of how to continue the procedure is similar to computing the
volume of a parallelogram spanning a pair of vectors $\v_1$ and $\v_2$:
first, pick one of them, say $\v_1$, and compute its length; then,
multiply that by the length of $\v_2$ computed along the direction
orthogonal to $\v_1$. An algorithm proposed by \cite{dpp-independence}
implements this probabilistically:
\begin{enumerate}
\item Sample one point $\v_i$ with $\Pr(i)\propto \|\v_i\|^2$;
\item Project all points $\v_j$ onto the subspace orthogonal to the sampled
  point $\v_i$;
\item Update the probabilities and go to step 1.
\end{enumerate}
This procedure will necessarily end after $k$ rounds, because all of the points will be projected onto the null space. 
At this point, we return the set of $k$ selected indices distributed as $\DPP(\U\U^\top)$. 
A straightforward implementation of this takes $O(nk^2)$ time.
However, \cite{minimax-experimental-design} showed that after computing the initial square norms $\|\v_i\|^2$, the algorithm can be implemented in time $O(k^3\log k)$ via rejection sampling. 

The overall sampling procedure suggested by \cite{dpp-independence} is very efficient, if we are given the eigendecomposition of kernel $\K$ or of the L-ensemble kernel $\L$. 
It can also be easily adapted to
sampling cardinality constrained DPPs. However, obtaining the
eigendecomposition itself can be a significant bottleneck: it
costs $O(n^3)$ time for a general $n\times n$ kernel. If we are
given a tall $n\times d$ matrix $\X$ such that $\L=\X\X^\top$, as was the
case in Section \ref{sxn:applications}, then the sampling cost can be
reduced to $O(nd^2)$ \cite{structured-dpp}. 
There have been a number of attempts
\cite{dpp-nystrom,dpp-coreset} at avoiding the eigendecomposition in
this procedure, leading to several approximate algorithms. Finally, approaches
using other factorizations of the kernel matrix have been proposed
\cite{efficient-volume-sampling,poulson2020}, and these offer
computational advantages in certain settings.

\subsection{DPPs: Intermediate sampling}
\label{s:intermediate}
The DPP sampling algorithms from Section \ref{s:mixture} can be
accelerated with a recently introduced technique
\cite{leveraged-volume-sampling,dpp-intermediate,dpp-sublinear}, 
which uses leverage score sampling to reduce the size of the $n\times n$
kernel matrix, without distorting the underlying DPP distribution. 
Recall from Section \ref{s:marginals} that  i.i.d.~leverage score
sampling can be viewed as an  
approximation of a DPP in which we ignore the negative correlations
between the sampled points. Naturally, in most cases such a sample as
a whole will be a very poor approximation of a DPP.
However, with high probability, it contains a DPP of a smaller size (Theorems~\ref{t:dlogd}~and~\ref{t:dsquared}). 

This motivates a strategy called \emph{distortion-free intermediate sampling}.
To explain how this strategy can be implemented, we will consider the special case of a Projection DPP. 
Using the notation from Section \ref{s:mixture}, let $S\sim\DPP(\U\U^\top)$, where $\U$ consists of $k$ orthogonal columns, and its rows are denoted $\v_i^\top$. 
\cite{leveraged-volume-sampling} showed that an i.i.d.~sample of
indices $i_1,...,i_t$ of size $t=O(k^2)$ drawn proportionally to the
leverage scores, i.e., $\Pr(i_t=j)\propto \|\v_j\|^2$, with high
probability contains a subset $S\sim\DPP(\U\U^\top)$ that can be found
efficiently, via the following procedure:
\begin{enumerate}
\item  
Sample $i_1,...,i_t$ i.i.d. $\propto(\|\v_1\|^2,...,\|\v_n\|^2)$;
\item 
Construct $\Ubt$ from the $t$ rows $\frac{1}{\|\v_{i_j}\|}\v_{i_j}^\top$;
\vspace{-2mm}
\item 
With prob.~$1-\det\!\big(\frac1t\Ubt^\top\Ubt\big)$ go back to 1;
\item 
Return $S = \{i_j\!: j\!\in\! \St\}$ for $\St \sim \DPP(\Ubt\Ubt^\dagger)$.
\end{enumerate}
This procedure essentially reduces the task of sampling from a DPP over a
large domain $\{1,...,n\}$ into
sampling from a potentially much smaller domain of size $O(k^2)$. In
particular, step 3 allows us to \emph{correct} the initial
i.i.d.~sample in such a way that the final result is exactly
$S\sim\DPP(\U\U^\top)$. 
This scheme can be extended
to any Projection DPP or L-ensemble (where ridge leverage scores are
used instead of the standard leverage scores), and neither the 
eigendecomposition nor the leverage scores need to be computed
exactly. \cite{dpp-sublinear} summarized this in the following result
for sampling general L-ensembles.
\begin{theorem}
Let $S_1,S_2$ be i.i.d.~random sets from $\DPP_L(\L)$, with $k=\E[|S|]$
or from any $k$-$\DPP(\L)$. Then, given access to $\L$, we can return
\begin{enumerate}
\item first, $S_1$ in:\quad $n\cdot\poly(k)\,\polylog(n)$ time,
\item then, $S_2$ in:\hspace{6mm} $\poly(k)$ time.
  \end{enumerate}
\end{theorem}

\noindent
Analogous time complexity statements can be provided when
$\L=\X\X^\top$. In this case, the first sample can be obtained in $O(nd\log n+
\poly(d))$ time, and each subsequent sample takes $\poly(d)$ time \cite{dpp-intermediate}.

\subsection{DPPs: Monte Carlo sampling}
\label{s:mcmc}
A completely different approach of (approximately) sampling from a
DPP was proposed by \cite{rayleigh-mcmc}, who
showed that a simple fast-mixing Monte Carlo Markov chain (MCMC) algorithm has a
cardinality constrained L-ensemble $k$-$\DPP_L(\L)$ as its stationary
distribution. The state 
space of this chain consists of subsets $S\subseteq[n]$ of some fixed
cardinality $k$. At each step, we choose an index $i\in S$ and
$j\not\in S$ uniformly at random. Letting
$T=S\cup\{j\}\backslash\{i\}$,  we transition from $S$ to $T$ with probability
\[\frac12\,\min\bigg\{1,\ \frac{\det(\L_{T,T})}{\det(\L_{S,S})}\bigg\},\]
and otherwise, stay in $S$. It is easy to see that the stationary
distribution of the above Markov chain is $k$-$\DPP_L(\L)$. Moreover,
\cite{rayleigh-mcmc} showed that the mixing time can be bounded as follows.
\begin{theorem}
The number of steps required to get to within $\epsilon$ total variation
distance from $k$-$\DPP_L(\L)$ is at most
$\poly(k)\,O\big(n\log(n/\epsilon)\big)$. 
\end{theorem}
The advantages of this sampling procedure over the algorithm of \cite{dpp-independence} are that we are not required to perform the eigendecomposition and that the computational cost of the MCMC chain scales linearly with $n$. 
The disadvantages are that the sampling is approximate and that we have to run the entire chain every time we wish to produce a new sample $S$.

In addition to the approach of \cite{rayleigh-mcmc} for $k$-DPPs,
\cite{gautier2017zonotope} proposed an MCMC designed specifically for Projection
DPPs, and another Markov chain was developed for unconstrained
L-ensembles by \cite{kdpp-mcmc}. Some of these MCMC approaches also
apply beyond DPPs, to \emph{Strongly
  Rayleigh} measures (see Figure \ref{f:dpp-diagram}). 

\section{Conclusions}
\label{sxn:conc}

We have briefly surveyed two established research areas which exhibit deep connections that have only recently began to emerge:
\begin{enumerate}
\item Randomized Numerical Linear Algebra; and
\item Determinantal Point Processes.
\end{enumerate}
In particular, we discussed recent developments in applying DPPs to classical tasks in RandNLA, such as least squares regression and low-rank approximation; and we surveyed recent results on sampling algorithms for DPPs, comparing and contrasting several different~approaches.

We expect that these connections will be fruitful more generally.
As an example of this, we briefly mention a recently proposed mathematical framework for studying determinants, which played a key role in obtaining some of these results. 

\noindent
\textbf{Determinant preserving random matrices.} 
A square random matrix $\A$ is determinant preserving (d.p.) if all of its
sub-determinants commute with taking the expectation, i.e., if:
\begin{align*}
  \E\big[\!\det(\A_{S,T})\big] =
  \det\!\big(\E[\A_{S,T}]\big),
\end{align*}
for all index subsets $S$, $T$ of the same size. Not all random
matrices satisfy this property (e.g., take $X\,\I_2$ for $X$ standard
Gaussian), however there are many non-trivial examples. Moreover, this class of
random matrices possesses a useful algebraic structure: if $\A$
and $\B$ are independent and d.p., 
then both $\A+\B$ and $\A\B$ are also determinant preserving. 
The first examples of d.p.~matrices where given by \cite{dpp-intermediate} (used in the analysis of a fast DPP sampling algorithm) and \cite{determinantal-averaging} (used for eliminating bias in distributed optimization), and further discussion can be found in~\cite{surrogate-design}.

Of course, our survey of the applications of DPPs necessarily excluded many areas where this family of distributions appears. 
Here, we briefly discuss some other applications of DPPs which are relevant in the context of numerical linear algebra and RandNLA but did not fit in the scope of this work. 

\noindent
\textbf{Implicit regularization.} \
In many optimization tasks (e.g., in machine learning), the true
minimizer of a desired objective is not unique or not computable
exactly, so that the choice of the optimization procedure affects the
output.  
Implicit regularization occurs when these algorithmic choices provide
an effect similar to explicitly introducing a regularization penalty
into the objective.  
This has been observed for approximate solutions returned by
stochastic and combinatorial optimization algorithms \cite{Mah12}, but
a precise characterization of this phenomenon for RandNLA sampling
methods has proven challenging.  
Recently, DPPs have been used to derive exact expressions for implicit
regularization in RandNLA algorithms \cite{surrogate-design,fanuel2020diversity},
connecting it to a phase transition called the double descent
curve \cite{BHMM19}.

\noindent
\textbf{Experimental design.} \
In statistics, the task of
  selecting a subset of data points for a down-stream regression task
  is referred to as experimental design~\cite{optimal-design-book}. In
  this context, it is often 
  assumed that the coefficients $y_i$ (or responses) are random
  variables obtained as a linear transformation of the vector $\x_i$
  distorted by some mean zero noise. A number of optimality criteria
  have been considered for selecting the best subsets in experimental
  design. DPP subset selection has been shown to provide useful
  guarantees for some of the most popular criteria (such as for A-optimality and
  D-optimality), leading to new approximation algorithms
  \cite{proportional-volume-sampling,symmetric-polynomials,bayesian-experimental-design}.

  \noindent
\textbf{Stochastic optimization.} \ 
Randomized selection of mini-batches of data or subsets of parameters
has been very successful in speeding up many iterative optimization
algorithms. Here, non-uniform sampling can be used to reduce the
variance in the iteration steps. In particular, \cite{dpp-minibatch}
showed that using a DPP for sampling mini-batches in stochastic
gradient descent improves the convergence rate of the optimizer,
whereas \cite{randomized-newton} used a DPP sampler for the Stochastic
Dual Newton Ascent method, showing an improved convergence analysis.

\noindent
\textbf{Monte Carlo integration.} \
DPPs have been shown to achieve theoretically improved
guarantees for numerical integration, i.e., using a weighted sum of
function evaluations to approximate an integral. In particular,
\cite{dpp-mcmc} constructed a DPP for which the root mean squared errors
of Monte Carlo integration decrease as $n^{-(1+1/d)/2}$, where $n$ is
the number of function evaluations and $d$ is the dimension.  This is
faster than the typical $n^{-1/2}$ rate. See
\cite{belhadji19,gautier19} for other results on Monte Carlo
integration with DPPs.

In conclusion, despite having been studied for at least forty five years, DPPs are enjoying an explosion of renewed interest, with novel applications emerging on a regular basis. 
Their rich connections to RandNLA, which we have only briefly summarized and which offer a nice example of how deep mathematics informs practical problems and vice versa, provide a particularly fertile ground for future work. 

\subsubsection*{Acknowledgements}

We would like to acknowledge DARPA, NSF (via the TRIPODS program), and
ONR (via the BRC on RandNLA) for providing partial support
for this~work.

\bibliographystyle{plain}
\renewcommand\refname{{\large References}}
{\footnotesize
\bibliography{pap}
}

\newpage
\appendix

\end{document}